\documentstyle[12pt,a4wide]{article}
\newcommand{\nin}{\noindent}
\newcommand{\be}{\begin{equation}}
\newcommand{\ee}{\end{equation}}
\newcommand{\bea}{\begin{eqnarray}}
\newcommand{\eea}{\end{eqnarray}}
\newcommand{\br}{\hskip .25cm/\hskip -.25cm}

\newcommand{\nonu}{\nonumber\\}

\newcommand{\dg}{^\dagger}
\newcommand{\ol}{\overline}

\begin{document}

\begin{titlepage} 

\begin{flushright} 
hep-ph/0111310
\end{flushright} 

\vspace{0.1in}

\begin{center}

{\Large {\bf Radiatively-induced 
Magnetic moment in four-dimensional 
anisotropic QED in an external magnetic field}}
\end{center}

\vspace{0.1in} 

\begin{center} 

{\bf Jean Alexandre} and {\bf Nick E. Mavromatos} \\

\vspace{0.1in} 
King's College London, Department of Physics - Theoretical Physics,
Strand, London WC2R 2LS, U.K. \\

\end{center} 

\vspace{0.2in} 

{\small We discuss one-loop radiatively-induced magnetic moment
in four-dimensional quantum electrodynamics (QED)
with anisotropic coupling, and examine various cases 
which may be of interest in effective gauge theories
of antiferromagnets, whose planar limit coresponds to 
highly anisotropic QED couplings. We find a different
scaling with the magnetic field intensity in case there 
are extra statistical gauge interactions in the model
with spontaneous symmetry breaking. Such a case is encountered
in the $CP^1$ $\sigma$-model sector of effective spin-charge
separated gauge theories of antiferromagnetic systems.
Our work provides therefore additional ways of 
possible experimental probing of the gauge nature 
of such systems.}

\end{titlepage}

\section{Introduction}

In a recent work~\cite{anisotrop} we have examined
dynamical mass generation for fermionic fields of
four-dimensional quantum electrodynamics (QED) 
with spatially anisotropic couplings across planar
two-dimensional surfaces in the $xy$ planes, in the presence of an external
homogeneous strong 
magnetic field along the $z$ direction of space. 
The phenomenon of magnetically induced 
fermionic mass 
is known as {\it magnetic catalysis}~\cite{gusynin}, and has wide applications
ranging from condensed matter~\cite{condmat} 
to early universe~\cite{universe}.   
It was found in \cite{anisotrop} that the magnetically 
induced mass depended on the anisotropy parameter, and it was maximum
for strong anisotropic (effectively planar QED) situations.
In the strong anisotropic limit, the induced mass looks as
if it is a 
parity-invariant three-dimensional mass among the effectively induced
three-dimensional fermion species on the plane, as a result of 
appropriate dimensional reduction of the four-dimensional spinors.  

The presence of an external magnetic field, however, breaks parity explicitly,
and this should be somehow seen in the effective theory.
Indeed, this is what happens as a result of the radiatively-induced 
magnetic moment of the fermions. At one (and higher) loop 
there will be induced a Pauli-type coupling $F_{\rm ext}^{\mu\nu}\sigma_{\mu\nu}$,
where $\sigma_{\mu\nu}=\frac{1}{2}[\gamma_\mu, \gamma_\nu ]$ in the 
effective one-loop lagrangian, where $F_{\rm ext}$ denotes the field
strength of the external gauge potential, corresponding to the 
applied magnetic field. 
{}From a physical point of view, such a situation may be encountered
in effective gauge theories of (doped) antiferromagnets, 
involving spin-charge separation \cite{anderson,gauge}, 
in the phase where the spin degrees of 
freedom (spinons) acquire a mass gap, and thus have been integrated 
out from the effective path-integral of the low-energy (massless) 
degrees of freedom, such as holons (fermion fields electrically charged). 
Indeed, in such a case, the low-energy effective lagrangian near the nodes
of the fermi surface of such systems, which are known to 
be $d$-wave superconductors, consists of relativistic electrically-charged
fermionic matter coupled to two-kinds of gauge fields: statistical ones, which 
is an effective way of describing the magnetic Heisenberg interactions
in the underlying microscopic condensed-matter model, and 
real electromagnetic ones. 

As we shall show below, the scaling 
of the induced magnetic moment with the magnetic-field intensity
is different in case there is spontaneous breaking in the 
statistical gauge sector,
compared to the situation where the gauge field is massless. 
This will allow one to discuss experimental probing of 
the nature of the gauge interactions in such systems, with 
obvious phenomenololgical significance.

The structure of the article is as follows: in section 2 we discuss
the anisotropic QED case setting up notations and conventions
for completeness. In section 3 we calculate
the one-loop induced 
magnetic moment for strong external fields, by resorting to 
lowest-Landau level computation, which is the only approximation 
that allows analytic treatment. We discuss quantum fluctuations
assuming a generic anisotropic coupling to the
statistical gauge field.
The computation is done for a massless 
gauge field as well as for a massive one (e.g. in case of 
spontaneous gauge symmetry breaking). 
The computation demonstrates clearly 
a different scaling of the radiatively-induced magnetic moment 
with the magnetic-field intensity between the two cases.
In section 4 we discuss a specific physical application of interest to
condensed matter physics, namely we present 
a
model 
where the spontaneous symmetry breaking occurs. 
The model is a four-dimensional $CP^{1}$ $\sigma$-model 
coupled to fermions, with a $SU(2)$ gauge symetry. 
We show that the integration over the 
bosonic field induces the non-Abelian kinetic term
for the gauge field. The model may be 
linked to 
the effective low-energy theory of doped antiferromagnets
in the spin-charge separated phase, and we discuss how
it is connected to the QED case
discussed in previous sections. Finally, conclusions and outlook
are presented in section 5. Technical aspects of our approach 
are presented in an Appendix.

\section{Anisotropic four-dimensional QED in an external magnetic field}

Our starting point 
will be the following four-dimensional space-time 
Lagrangian density, which includes the anisotropy~\cite{anisotrop}:

\be\label{modelanisotrop}
{\cal L}=-\frac{1}{4}F_{\mu\nu}F^{\mu\nu}+\ol\psi
\left[i\br\partial-g\br A-x(i\partial_3-eA_3)\gamma^3\right]\psi,
\ee

\nin where $\mu,\nu= 0, \dots 3$, and 
the parameter $x$ controls the anisotropy: $x=0$ is the usual 
isotropic QED
and $x=1$ the completely anisotropic case, i.e. the fermions live 
effectivelly in 2+1 dimensions,
whereas the gauge field still lives in 3+1 dimensions.
In the absence of an external field, the bare fermion propagator is
(the terminology `bare' is intended to imply quantities  
without interaction with the dynamical field) 
\be
iS^{-1}(p)=\br p-xp_3\gamma^3-m,
\ee

\nin and the bare vertices

\bea
\Lambda^\mu&=&\gamma^\mu~~~~\mbox{if}~~\mu\ne 3\nonu
\Lambda^3&=&(1-x)\gamma^3.
\eea

\nin
In the presence of a constant external field $A^{ext}_\mu$, 
it is known \cite{schwinger} that the bare
fermion propagator 
acquires a phase and becomes

\be\label{separ}
S(y,z)=e^{iey^\mu A^{ext}_\mu(z)}\tilde S(y-z),
\ee

\nin which has also been shown to be true via 
a non perturbative method for the full dressed
propagator \cite{ext}. In the propagator (\ref{separ}), the coupling $e$ to the electromagnetic 
field is different from the coupling $g$ to the statistical gauge field which represents 
an effective interaction in the condensed-matter system, as will be discussed in section 4.

In the case of a constant magnetic field, we will use the so-called lowest Landau level (LLL)
approximation \cite{gusvertex} valid for strong fields. 
In this approximation the Fourier transform of the
translational invariant part $\tilde S$ of the fermion propagator is
  
\be\label{SL}
\tilde S^L(p)=e^{-p_\bot^2/|eB|}\frac{i(1-i\gamma^1\gamma^2)}{p_0\gamma^0+(1-x)p_3\gamma^3-m},
\ee

\nin where the magnetic field is in the direction 3 and $\bot$ denotes the transverse
directions 1 and 2 in the plane where the fermions are localized in the 
fully anisotropic case~\cite{anisotrop}. 

The one-loop vertex was computed in the isotropic case \cite{gusvertex} and its general
expression is

\bea\label{vertex}
\Gamma^\lambda(r;y,z)&=&-e^2\Lambda^\mu S(y,r)\Lambda^\lambda S(r,z)\Lambda^\nu D_{\mu\nu}(z-y)\nonu
&=&e^{iey^\mu A^{ext}_\mu(z)}\tilde \Gamma^\lambda(y-r,z-r).
\eea

\nin We note that the expression (\ref{vertex}) involves the Green functions in
coordinate space, which is the necessary starting point so as to take into account
the phase factors that appear in Eq.(\ref{separ}).
The Fourier transform of the one-loop vertex is then

\bea\label{oneloopvertex}
\tilde \Gamma^\lambda(k_1,k_2)&=&-g^2\int_{pq}\int_ze^{iz(p-q-k_2)}
D_{\mu\nu}(q)\nonu
&&\times\Lambda^\mu \tilde S(k_1+eA^{ext}(z)-q)
\Lambda^\lambda \tilde S(p)\Lambda^\nu,
\eea

\nin where for simplicity we used the notation 
$\int_p=\int d^4p/(2\pi)^4$ and $\int_z=\int d^4z$.

\section{Radiatively-induced magnetic moment}

In this section, we compute the 
anomalous magnetic moment, induced radiatively at one loop
in the model of the previous section.
We assume  the following configuration for the 
external gauge field : $A^{ext}(z)=(0,-Bz_2/2,Bz_1/2,0)$, 
which corresponds to the 
physically relevant case of a constant magnetic field along the 
$z$ axis (direction 3).  

Consider the full vertex function  
in isotropic four-dimensional QED without external field~\cite{zuber}:

\be 
\Gamma^\lambda(0,k)=\gamma^\lambda F_1(k)+\frac{i}{2m}\sigma^{\lambda\rho}k_\rho F_2(k).
\ee
The anomalous magnetic moment $\mu_0$ of the fermions 
is given by $F_2(0)$.

We now note that, in the presence of an external magnetic field, 
the corrections along the transverse directions 1,2, 
$\Gamma^\bot$ are zero in the 
LLL approximation \cite{gusvertex}. To see this, 
let us introduce the projectors 

\be
P^{(\pm)}=(1\pm i\gamma^1\gamma^2)/\sqrt{2}.
\ee

\nin Because we have
$P^{(-)}\gamma^\bot=\gamma^\bot P^{(+)}$ and $P^{(-)}P^{(+)}=0$, the components $\Gamma^\bot$ 
must vanish since $\tilde S(p)$ is proportional to $P^{(-)}$. 

To determine the induced magnetic moment, 
we will then look for the behaviour of the component $\Gamma^3$ and consider
its part which is proportional to $\sigma^{30}k_0=\gamma^3\gamma^0k_0$.
For the photon propagator, we will take the Feynman gauge 
and we introduce the Euclidean $\gamma$ matrices $\gamma^\mu$, $\mu=1,...,4$ satisfying 
$\{\gamma^\mu,\gamma^\nu\}=-2\eta^{\mu\nu}$.

Starting from (\ref{oneloopvertex}), the
integration over $z_\|$ can be performed and leads to a $\delta^2(q_\|-p_\|+k_\|)$. 
The integrations over $p_\bot$ and $z_\bot$ are Gaussian and give

\bea\label{vertexint}
\tilde\Gamma^3(0,k)&=&m\Lambda^\mu\Lambda^3\sqrt{2}P^{(-)}\Lambda^4\Lambda_\mu\nonu
&\times&g^2\int_{p_\|q_\bot}e^{-\frac{1}{|eB|}\left(q_\bot^2+\frac{1}{2}k_\bot^2
+q_\bot k_\bot +i(q_2k_1-q_1k_2)\right)}\nonu
&\times&\frac{1}{(p_4-k_4)^2+(1-x)^2(p_3-k_3)^2+m^2}\nonu
&\times&\frac{1}{p_4^2+(1-x)^2p_3^2+m^2}\nonu
&\times&\frac{1}{q_\bot^2+(p_\|-k_\|)^2}+\mbox{other terms}
\eea

\nin where `other terms' denote the terms which are not proportional to the
expected $\Lambda^3\Lambda^4$ contribution obtained here as:

\be
\Lambda^\mu\Lambda^3\sqrt{2}P^{(-)}\Lambda^4\Lambda_\mu=
(x^2+2x-2)\Lambda^3\Lambda^4\sqrt{2}P^{(-)}+
2\Lambda^3\Lambda^4\sqrt{2}P^{(+)},
\ee

\nin We finally define the anomalous magnetic moment $\mu_B$ by the expression

\be
\tilde\Gamma^3(0,k)=\sqrt{2}P^{(+)} \frac{\mu_B}{2m}\Lambda^3\Lambda^4 k_4 +{\cal O}(k^2)
+\mbox{other terms}.
\ee

\nin After integration over $p_\|$ in (\ref{vertexint}) we find: 

\be\label{magmom}
\mu_B(x)=\frac{g^2}{4\pi^2}\int_0^1 
\frac{d\eta~\eta(1-\eta)}{\sqrt{\eta+(1-\eta)(1-x)^2}}
\int_0^\infty  
\frac{du~e^{-u\frac{m^2}{|eB|}}}{[1+\eta(u-1)]^2},
\ee

\nin where $\eta$ is a Feynman parameter.
The integration on $u$ can be approximated in the following
way for $|eB|>>m^2$:

\be\label{approx}
\int_0^\infty \frac{du~e^{-u\frac{m^2}{|eB|}}}{[1+\eta(u-1)]^2}\simeq
\int_0^\frac{|eB|}{m^2} \frac{du}{[1+\eta(u-1)]^2}\simeq
\frac{1}{\eta(1-\eta)}, 
\ee

\nin such that the integration over $\eta$ in Eq.(\ref{magmom}) gives finally

\be\label{magmomfinal}
\mu_B(x)\simeq 
\frac{g^2}{2\pi^2}\frac{1}{2-x}.
\ee

\nin Let us discuss Eq.(\ref{magmomfinal}). The first 
new result is the following: in the isotropic
case, $x=0$, the anomalous magnetic moment $\mu_B(0)$ is twice 
the well-known result 
obtained in the absence of magnetic field~\cite{zuber}: 
$\mu_0=g^2/8\pi^2$. Then it increases with the anisotropy,
and reaches its maximum value in the fully anisotropic regime, $x=1$, 
which  is again twice its isotropic value, in other words, four times
the value without an external field, $4\mu_0$.  

Thus the anisotropy
has the same impact on the anomalous magnetic moment than 
it had on the dynamical mass \cite{anisotrop},
i.e. it provides an enhancement, which however, in this case 
is not exponential as it was for the
dynamical mass. 
Note that $\mu_B(x)$ does not depend on the fermion mass in 
the LLL approximation. We will study now the case of a 
massive gauge field, where it will be shown that the anomalous
magnetic moment depends 
on the fermion dynamical mass, and therefore will 
imply a more complicated scaling 
with the magnetic field, given the associated 
magnetic-field dependence of the fermion mass~\cite{anisotrop}. 

Let us suppose that the gauge field has acquired 
a mass $M$ via some Higgs mechanism that we will discuss
in the next section. In this case, Eq.(\ref{magmom}) becomes

\be
\mu_B(x)=\frac{g^2}{4\pi^2}\int_0^1
\frac{d\eta~\eta(1-\eta)}{\sqrt{\eta+(1-\eta)(1-x)^2}}
\int_0^\infty
\frac{du~e^{-u\frac{m^2}{|eB|}}}{[1+\eta(u+\epsilon\phi^2)]^2},
\ee
  
\nin where $\phi^2=|(M/m)^2-1|$ and $\epsilon=\mbox{sign}\{(M/m)^2-1\}$. 
The approximation (\ref{approx}) then gives:

\be
\mu_B(x)\simeq\frac{g^2}{4\pi^2}\int_0^1
\frac{d\eta~(1-\eta)}{(1+\epsilon\eta\phi^2)\sqrt{\eta+(1-\eta)(1-x)^2}}.
\ee

\nin We are interested in the anisotropic case $x=1$. The integration over $\eta$ gives then

\bea\label{finalM}
\mu_B(1)&=&\frac{g^2}{2\pi^2}\left[\frac{\tan^{-1}\phi}{\phi}\left(1+\frac{1}{\phi^2}\right)
-\frac{1}{\phi^2}\right]~~\mbox{if}~M\ge m\nonu
\mu_B(1)&=&\frac{g^2}{2\pi^2}\left[\frac{\tanh^{-1}\phi}{\phi}\left(1-\frac{1}{\phi^2}\right)
+\frac{1}{\phi^2}\right]~~\mbox{if}~M\le m
\eea

\nin where $\tan^{-1}(z)=\arctan(z)$ and $\tanh^{-1}(z)=1/2\ln\left((1+z)/(1-z)\right)$.
When $M<<m$, we recover the result (\ref{magmomfinal}) from Eq.(\ref{finalM}) since

\be\label{masslessmom} 
\mu_B(1)\to \frac{g^2}{2\pi^2}~~\mbox{when}~\phi\to 1.
\ee

\nin In the other limit, when $m\to 0$ or $\phi\to\infty$, we also recover the 
fact that $\mu_B$ vanishes. 

The magnetic field dependence becomes non-trivial 
in the case where $M>>m$. 
Our motivation to study this case is given in the next section. 
$m$ is the magnetically induced (dynamically generated) 
fermion mass gap~\cite{anisotrop} 

\be\label{dynx1}
m_{dyn}(x=1)\simeq\frac{g^2}{4\pi}\sqrt{|eB|}.
\ee

\nin Furthermore, as will be discussed in the next section, 
$M$ does not depend on the magnetic field.
Therefore we have:  

\be\label{magmommass}
\mu_B(1)\simeq\frac{g^4\sqrt{|eB|}}{8\pi^3M} ~~\mbox{when}~\phi>>1,
\ee

\nin i.e. the anomalous magnetic moment increases with the applied field. 
We stress that this result is valid 
as long as $m<<M$, which 
can be achieved in a wide region of magnetic field, due to 
the relation (\ref{dynx1}). 

The results in this section, then, may be summarized as follows:
in the case of a massive
gauge field the anomalous magnetic 
moment scales as the square root of $|eB|$, whereas
it is independent of the magnetic field if the 
gauge field is massless.

\section{An application: spin-charge separating effective theories of 
antiferromagnets} 

In this section we discuss a physically interesting 
potential application of the 
above phenomenon, of relevance to condensed-matter physics. 
Namely, we shall discuss
a (continuum) model related to the 
low-energy physics of planar antiferromagnets in the spin-charge
separated phase. We shall link this model with the 
case of the anomalous-magnetic moment induced in the 
spontaneously broken gauge symmetry situation (\ref{magmommass})
encountered in the previous section.

In the condensed-matter model of doped antiferromagnets 
discussed in ref. 
\cite{farakos}, we made use of an
(approximate for low-doping) particle-hole symmetry 
in effective microscopic models of doped antiferromagnets,
to arrive at the following 
spin-charge separation ansatz which had
a manifest $SU(2)$ local gauge symmetry 
even away from half-filling (doped case) 

\begin{equation} 
\left(\begin{array}{cc} c_1 \qquad c_2 \\ c_2^\dagger \qquad 
-c_1^\dagger \end{array} \right) = \left(\begin{array}{cc} \psi_1 
\qquad \psi_2 \\ \psi_2^\dagger 
\qquad -\psi_1^\dagger \end{array}\right) 
\left(\begin{array}{cc} z_1 \qquad -\overline{z}_2 \\
z_2 \qquad \overline{z}_1 \end{array}\right)
\label{ansatz} 
\end{equation}

\nin Here, $c_{\alpha}$, with $\alpha \in\{1,2\}$ 
are electron operators with spin up or down, and 
$\psi_\alpha$, $z_\alpha$ with $\alpha\in\{1,2\}$ are fermions
(Grassmann) and bosons respectively.  The $\psi_\alpha$ describe 
elelctrically-charged degrees of
freedom (holons) and the $z_\alpha$ describe the spin degrees of freedom
(magnons) .  The
index $\alpha$ is related to the underlying bipartite
(antiferromagnetic) lattice structure.

In \cite{farakos} we have restricted ourselves to the planar case
for such a separation, in which all the degrees of freedom are
confined on the $xy$ plane. 
It is {\it only in that case} that the 
fermionic matrix 
in (\ref{ansatz}) can be transformed,
at a continuum effective low-energy lagrangian level, 
to appropriate 
Nambu-Dirac two-spatial-dimensional spinors $\Psi _c$,
where $c =1,2$ is a colour $SU(2)$ index of
the fundamental representation of $SU(2)$. 
The effective lagrangian describes nodal excitations
around zeroes of the fermi surface of the microscopic model. 

For our purposes here we shall {\it assume}  
that a local $SU(2)$ symmetry also characterizes 
a fully four-dimensional case of nodal excitations, in which 
an appropriately modified spin-charge separation 
(\ref{ansatz}) is valid,
but the $SU(2)$ gauge coupling to be spatially anisotropic.
The effective planar case, then, corresponds to the highly anisotropic
limit $x=1$ for the coupling. 
In that case, the electrically-charged 
nodal excitations are viewed as fully-fledged
four-dimensiomal {\it relativistic} Dirac spinors,
while the spin (boson) parts retains its $CP^{1}$ $\sigma$-model
form. The assumed {\it continuum} 
effective lagrangian of the nodal liquid then of 
excitations around the fermi-surface nodes of this four-dimensional
problem has the form: 

\bea\label{model} 
{\cal L}_{\rm eff}&=&\frac{1}{\gamma}\left|(\partial_\mu-ig{\cal A}_\mu)z\right|^2
+\lambda(|z|^2-M_z^2)\nonu
&&-e\ol\psi\br A^{ext}\psi
+\ol\psi\left[i\br\partial-g\br{\cal A}-x(i\partial_3-g{\cal A}_3)\gamma^3\right]\psi,
\eea 

\nin where $\gamma$ is a dimensionless coupling 
constant (in four space-time dimensions), and $|z|^2=z\dg z$. 
In the Lagrangian (\ref{model}),
${\cal A}_\mu=a_\mu^c{\cal T}^c$, $c=1,2,3$, and $a_\mu^c$ are the
gauge bosons of the statistical gauge group $SU(2)$ with generators ${\cal T}^c$. 
$\lambda$ is a Lagrange multiplier field implementing the $CP^{1}$ constraint $|z|^2=M_z^2$
We did not write the anisotropic coupling of the holons to the external field
since the latter has no component in the direction 3. Finally, our model also takes into account
the spin-charge separation at the level of the coupling to the statistical gauge field, since only
the holons are coupled anisotropically whereas the spinons are coupled isotropically. This is a 
consequence of different hopping of the holons and spinons between the antiferromagnetic planes.

We will now discuss the emergence of a {\it dynamical}, and {\it massive},
$SU(2)$ gauge field, after integrating out the $z$ degrees of freedom.
We first assume that the field $z$ has a non zero expectation value

\be\label{vevz} 
< z > = z_0 \ne 0
\ee 

\nin where $z_0$ is assumed constant to a first approximation,
spatial inhomogeneities are suppressed for our purposes here. 
The bosonic contribution to the Lagrangian (\ref{model}) reads then

\bea\label{translation}
&&\frac{1}{\gamma}\left|(\partial_\mu-ig{\cal A}_\mu)z\right|^2+\lambda(|z|^2
-M_z^2)\nonu
&&=\frac{g^2}{\gamma}z_0\dg{\cal A}_\mu{\cal A}^\mu z_0
+\frac{1}{\gamma}\left|(\partial_\mu-ig{\cal A}_\mu)\tilde z\right|^2
+\lambda|\tilde z|^2\nonu
&&+2{\cal R}e\left(z_0\dg\lambda\tilde z 
+i\frac{g}{\gamma}z_0\dg{\cal A}_\mu(\partial^\mu-ig{\cal A}^\mu)\tilde z\right)
+\lambda(|z_0|^2-M_z^2),
\eea

\nin where we used ${\cal A}_\mu\dg={\cal A}_\mu$. 
In Eq.(\ref{translation}), we see the appearence of a mass term for the three gauge fields.
In order of magnitude, the gauge-boson mass is

\be
M^2\simeq\frac{g^2}{\gamma}z_0\dg z_0,
\ee

\nin i.e. it is linked to the expectation value of the spinon field, as is usual in a
Higgs mechanism. Since the spinon field is neutral, i.e. not coupled to the electromagnetic
field, $M$ 
will be independent of the magnetic field, depending  only on the 
vacuum expectation value of $z$ only.

Some comments on the order of magnitude of $M$, as compared 
to the dynamical fermion mass $m$ are in order. 
In the physical situation we have in mind, 
the relativistic fermions encountered in the model 
represent the continuum limit of 
excitations of a microscopic coondensed-matter 
system near the nodes of a $d$-wave superconducting gap.
Experimentally, the disappearance of the nodes, and therefore the 
opening up of a gap for the quasiparticle excitations at those 
points on the fermi surface,
is observed
indirectly~\cite{krishana}, 
through plateaux in the thermal conductivity of the 
high-temperature $d$-wave superconducting materials,
below a certain temperature.
The later 
is much lower
than the bulk critical temperature of the superconductor.

In certain models~\cite{farakos,condmat},  
the magnetically induced fermion (holon) 
mass gap $m$ 
is found to be much smaller than the spinon gap $M_z$ 
and the 
gauge boson mass gap $M$, both related to spin 
degrees of freedom in the problem. In such cases one has 
$M \gg m$, a situation encountered in section 3, which implies
a non-trivial scaling (\ref{magmommass}) 
of the induced magnetic moment 
with the external field intensity.  Hence, 
by measuring experimentally such a 
scaling,
one can probe deeper into the  
possible gauge structure 
of such spin-charge separated systems.  
Recall from our discussion in section 3, that 
in the opposite situation, where 
the gauge field mass is smaller than the holon mass, there will be no 
appreciable scaling (\ref{masslessmom}) 
of the induced magnetic moment with the external field.

The gauge kinetic term will be obtained 
via the integration over the field $\tilde z$.
This integration was done in 2 dimensions in the Abelian
case \cite{polyakov}, without spontanous symmetry breaking for 
the field $z$. We show in the appendix 
that one can also recover the gauge kinetic term in 4 
dimensions, with a non-Abelian gauge symmetry, taking 
into account the non-vanishing vacuum expectation value of $z$. 
The path integration over $z$ fields, then,  
yields the folowing gauge-field contributions to the effective 
lagrangian: 

\be
{\cal L}_{\rm gauge}=-\frac{g^2}{96\pi^2\varepsilon} F_{\mu\nu}^a F^{a\mu\nu}
+\frac{g^2}{\gamma} z_0\dg {\cal A}_\mu {\cal A}^\mu z_0,
\ee

\nin where we used a dimensional regularization ($d=4-\varepsilon$) and 
$F_{\mu\nu}^a=\partial_\mu A_\nu^a-\partial_\nu A_\mu^a+gf^{abc}A_\mu^b A_\nu^c$.
In the framework of the 
condensed-matter effective study that we make, the pole $1/\varepsilon$ is understood as a 
natural U.V. cut-off of the theory, 
depending on the specific microscopic lattice system considered (e.g. 
$1/\varepsilon \sim {\rm ln}(L/a)$, where $L$ is an I.R. scale, and 
$a$ is an U.V. one).

Thus, starting from the Lagrangian (\ref{model}), we have 
derived the effective theory containing
holons, which in turn are 
anisotropically coupled to a massive dynamical $SU(2)$ gauge field.
In this way we have arrived at the situations 
described in section 3.

\section{Conclusion}

In this paper we have derived the radiatively-induced 
magnetic moment in the case of a four-dimensional 
gauge field theory model with 
an anisotropic coupling of the fermions to the gauge fields. 
We saw that the resulting
anomalous magnetic moment is independent 
of the magnetic field if the gauge field
is massless, whereas it scales as the square root of the 
magnetic field if the
gauge field acquires a mass via some (spontaneous gauge symmetry breaking) 
mechanism. 
This situation has to be distinguished 
from that discussed in 
\cite{momen}, where the anomalous magnetic moment in the presence of an
external 
magnetic field was computed in {\it strictly} 2+1 dimensions, and therefore 
did not lead to 
the same scaling. We believe that the present 
anisotropic situation is more suitable 
for the effective description of 
the planar high-temperature 
superconductors, which notably are quasi three-dimensional
systems, involving a 
small but finite electron hopping across the 
superconducting planes. 

The scaling of the induced magnetic moment with 
the magnetic field intensity constitutes 
another interesting experimental probe  
of the spin-charge 
seperation ansatz and the presence 
of spontanous symmetry breaking for the spinon
degrees of freedom, in the way explained in this article. 
The associated parity violation of the effective theory, 
which accompanies the apperance of
the external magnetic field, and manifests itself through 
the induced magnetic moment, may result in edge 
(parity-violating) currents
in the superconducting materials, 
whose intensity would depend on the applied magnetic field.
Such effects  
should be directly measureable.
We plan to return to a systematic analysis of such tests
in a forthcoming publication.

\section*{Aknowledgements} 

This work is supported by the Leverhulme Trust, UK.

\section*{Appendix: Generation of a 
dynamical gauge field in a four-dimensional $CP^1$ $\sigma$-model}

The purpose of this appendix is to perform explicitly, and discuss 
techincal details of, 
the path integration over the field $z$ in the Lagrangian 
(\ref{model}), so as to generate dynamically the kinetic term 
of the statistical gauge field.
 
In this integration, the linear terms in $\tilde z$ can be omitted in Eq.(\ref{translation}) for the 
following reason: let us consider the integral

\bea
&&\int dz \exp\left(-az-f(z^2)\right)\nonu
&=&\int dz \cosh(az)\exp\left(-f(z^2)\right)\nonu
&=&\int dz \exp\left(\ln\cosh(az)-f(z^2)\right)\nonu
&=&\int dz \exp\left(-f(z^2)+\frac{(az)^2}{2}+...\right),
\eea

\nin such that the term which was originally linear in 
$\tilde z$ actually appears only with even powers and therefore
has irrelevant contributions, since its square has already mass dimension 
5 (the field $\lambda$ has mass dimension 2).  
We have

\bea
&&\int{\cal D}[z]{\cal D}[z\dg]
\exp\left(-\int\left[\frac{1}{\gamma}\left|(\partial_\mu-ig{\cal A}_\mu)z\right|^2
+\lambda(|z|^2-m^2)\right]\right)\nonu
&=&\mbox{cte}\times\exp\left(-\int \frac{g^2}{\gamma}z_0\dg{\cal A}_\mu{\cal A}^\mu z_0\right)\nonu
&&\times\int{\cal D}[\tilde z]{\cal D}[\tilde z\dg]
\exp\left(-\int_{pq}\tilde z\dg(p){\cal O}(p,q)\tilde z(-q)+\mbox{irrelevant}\right),
\eea

\nin where the operator ${\cal O}$ is 

\bea
{\cal O}(p,q)&=&(\tilde\lambda+p^2)\delta(p+q)-(p-q)^\mu g{\cal A}_\mu(p+q)\nonu
&&+g^2\int_k{\cal A}_\mu(k){\cal A}^\mu(p+q-k)
\eea

\nin and $\tilde\lambda=\gamma\lambda$. The integration over $\tilde z$ leads to
the trace of the logarithm of ${\cal O}$ that we expand 
up to the forth order in the gauge field so as to recover
the non-Abelian gauge kinetic term:

\bea\label{tracelog}
&&\mbox{Tr} \ln\left[\frac{{\cal O}(p,q)}{\tilde\lambda+p^2}\right]\nonu
&=&g^2\mbox{tr}\int_k \Pi^{\mu\nu}_{(2)}(k){\cal A}_\mu(k){\cal A}_\nu(-k)\nonu
&+&g^3\mbox{tr}\int_{kp} \Pi^{\mu\nu\rho}_{(3)}(k,p){\cal A}_\mu(k){\cal A}_\nu(p){\cal A}_\rho(-k-p)\nonu
&+&g^4\mbox{tr}\int_{kpq} \Pi^{\mu\nu\rho\sigma}_{(4)}(k,p,q){\cal A}_\mu(k){\cal A}_\nu(p){\cal A}_\rho(q)
{\cal A}_\sigma(-k-p-q)\nonu
&+&\mbox{higher orders}, 
\eea

\nin where `Tr' 
denotes the trace over momenta and indices of $SU(2)$ generators, and 

\bea\label{tensors}
\Pi^{\mu\nu}_{(2)}(k)&=&\int_r\frac{1}{\tilde\lambda+r^2}\left(g^{\mu\nu}
-\frac{1}{2}\frac{(2r-k)^\mu(2r-k)^\nu}{\tilde\lambda+(k-r)^2}\right)\\
\Pi^{\mu\nu\rho}_{(3)}(k,p)&=&
\int_r\frac{p^\mu-2r^\mu}{[\tilde\lambda+(p-r)^2](\tilde\lambda+r^2)}\nonu
&&\times\left(g^{\nu\rho}+\frac{1}{3}(2r^\nu+k^\nu)
\frac{p^\rho-k^\rho-2r^\rho}{\tilde\lambda+(k+r)^2}\right)\nonu
\Pi^{\mu\nu\rho\sigma}_{(4)}(k,p,q)&=&\int_r\frac{1}{[\tilde\lambda+(p+k-r)^2][\tilde\lambda+r^2]}\nonu
&&\times\left(-\frac{1}{2}g^{\mu\nu}g^{\rho\sigma}+g^{\rho\sigma}(p^\nu-2r^\nu)
\frac{k^\mu+2p^\mu-2r^\mu}{\tilde\lambda+(p-r)^2}\right.\nonu
&&\left.~~+\frac{1}{4}(p^\nu-2r^\nu)(q^\rho+2r^\rho)
\frac{(k^\mu+2p^\mu-2r^\mu)(k^\sigma+p^\sigma-q^\sigma-2r^\sigma)}
{[\tilde\lambda+(p-r)^2][\tilde\lambda+(q+r)^2]}\right).\nonumber
\eea

\nin The integrals (\ref{tensors}) are computed within dimensional 
regularization ($d=4-\varepsilon$), and we obtain

\bea
\Pi^{\mu\nu}_{(2)}(k)&=&\frac{1}{48\pi^2\varepsilon}(g^{\mu\nu}k^2-k^\mu k^\nu)+\mbox{finite}\nonu
\Pi^{\mu\nu\rho}_{(3)}(k,p)&=&\frac{1}{72\pi^2\varepsilon}\left[
g^{\mu\rho}(2p^\nu-k^\nu)-g^{\nu\rho}(p^\mu+2k^\mu)+g^{\mu\nu}(k^\rho-p^\rho)
\right]+\mbox{finite}\nonu
\Pi^{\mu\nu\rho\sigma}_{(4)}(k,p,q)&=&\frac{1}{96\pi^2\varepsilon}\left[
2g^{\mu\nu}g^{\rho\sigma}-g^{\mu\rho}g^{\nu\sigma}-g^{\mu\sigma}g^{\nu\rho}\right]+\mbox{finite}
\eea

\nin We use then the following traces

\bea
\mbox{tr}\left({\cal T}^a{\cal T}^b\right)&=&\delta^{ab}\nonu
\mbox{tr}\left({\cal T}^a\left\{{\cal T}^b,{\cal T}^c\right\}\right)&=&0\nonu
\mbox{tr}\left({\cal T}^a\left[{\cal T}^b,{\cal T}^c\right]\right)&=&if^{abc}\nonu
\mbox{tr}\left(\left[{\cal T}^a,{\cal T}^b\right]\left[{\cal T}^c,{\cal T}^d\right]\right)&=&
-f^{abe}f^{cde}
\eea

\nin where $f^{abc}$ are the $SU(2)$ structure constants, and we find

\bea
\mbox{quadratic term:}&&\mbox{tr}\int_k \Pi^{\mu\nu}_{(2)}(k){\cal A}_\mu(k){\cal A}_\nu(-k)\\
&=&-\frac{g^2}{48\pi^2\varepsilon}\int_k (g^{\mu\nu}k^2-k^\mu k^\nu)A_\mu^a(k)A_\nu^a(-k)\nonu
&=&-\frac{g^2}{96\pi^2\varepsilon}\int_z (\partial_\mu A_\nu^a(z)-\partial_\nu A_\mu^a(z))
(\partial^\mu A^{\nu a}(z)-\partial^\nu A^{\mu a}(z))\nonu
\mbox{cubic term:}&&\mbox{tr}\int_{kp} \Pi^{\mu\nu\rho}_{(3)}(k,p){\cal A}_\mu(k){\cal A}_\nu(p){\cal A}_\rho(-k-p)\nonu
&=&-\frac{g^3}{72\pi^2\varepsilon}\int_{kp}3if^{abc}p^\mu A_\mu^a(k)A_\nu^b(p)A^{\nu c}(-k-p)\nonu
&=&-\frac{g^3}{48\pi^2\varepsilon}\int_{kp}if^{abc}A_\mu^a(k)A_\nu^c(-k-p)
\left(p^\mu A^{\nu b}(p)-p^\nu A^{\mu b}(p)\right)\nonu
&=&-\frac{g^3}{96\pi^2\varepsilon}\int_z 2f^{abc}A_\mu^a(z) A_\nu^c(z) 
\left(\partial^\mu A^{\nu b}(z)-\partial^\nu A^{\mu b}(z)\right)\nonu
\mbox{quartic term:}&&\mbox{tr}\int_{kpq} \Pi^{\mu\nu\rho\sigma}_{(4)}(k,p,q){\cal A}_\mu(k){\cal A}_\nu(p){\cal A}_\rho(q)
{\cal A}_\sigma(-k-p-q)\nonu
&=&-\frac{g^4}{96\pi^2\varepsilon}\int_{kpq} f^{abc}f^{ade}A_\mu^b(k) A_\nu^c(p) A^{\mu d}(q) A^{\nu e}(-k-p-q)\nonu
&=&-\frac{g^4}{96\pi^2\varepsilon}\int_z f^{abc}f^{ade}A_\mu^b(z) A_\nu^c(z) A^{\mu d}(z) A^{\nu e}(z)\nonumber
\eea

\nin such that the integration over $\tilde z$ gives finally the expected kinetic term 

\be\label{intfinal}
\mbox{Tr}\ln\left[\frac{{\cal O}(p,q)}{\tilde\lambda+p^2}\right]
=-\frac{g^2}{96\pi^2\varepsilon}\int_z F_{\mu\nu}^a F^{a\mu\nu}+~\mbox{higher orders},
\ee

\nin where $F_{\mu\nu}^a=\partial_\mu A_\nu^a-\partial_\nu A_\mu^a+gf^{abc}A_\mu^b A_\nu^c$.
In Eq.(\ref{intfinal}), `higher orders' 
denotes the irrelevant operators (mass dimension greater 
than 5) that are not taken into account in the low-energy derivative
expansion we study here. 
We note that the field $\tilde\lambda$ does not appear in Eq.(\ref{intfinal}) (it actually
only appears in the irrelevant terms) and thus its
integration can be omitted in the final path integral defining the effective model.

\end{document}